\documentclass[%
reprint,
 amsmath,amssymb,
 aps,
prb,
]{revtex4-2}

\usepackage{graphicx}% Include figure files
\usepackage{dcolumn}% Align table columns on decimal point
\usepackage{bm}% bold math
\usepackage{float}
\usepackage[colorlinks,
            urlcolor=blue,
            linkcolor=blue,
           anchorcolor=blue,
            citecolor=blue]{hyperref}
\begin{document}

\preprint{APS/123-QED}

\title{Acoustic helical dichroism in a one-dimensional lattice of chiral resonators}% Force line breaks with \\

\author{Qing Tong\textsuperscript{1}}
\author{Shubo Wang\textsuperscript{1,}}\thanks{Corresponding author: S.W. (shubwang@cityu.edu.hk)}
\affiliation{\textsuperscript{1}Department of Physics, City University of Hong Kong, Tat Chee Avenue, Kowloon, Hong Kong, China}%

\date{\today}% It is always \today, today,
            
\begin{abstract}
Circular dichroism and helical dichroism are intriguing chiroptical phenomena with broad applications in optical sensing and imaging. Here, we generalize one of the phenomena—helical dichroism—to acoustics. We show that a one-dimensional lattice of chiral resonators with loss can induce differential absorption of helical sounds (i.e. acoustic vortices) carrying opposite orbital angular momentum (OAM). This acoustic helical dichroism strongly depends on the rotation symmetry of the chiral resonators. A breaking of the $C_4$ rotation symmetry can induce coupling between the opposite chiral dipole modes of the resonators. This leads to OAM bandgaps and non-Hermitian exceptional points near the Brillouin-zone center and boundaries, which together give rise to significantly enhanced helical dichroism. The underlying physics can be well captured by an effective Hamiltonian that quantitatively reproduces the complex band structures. The acoustic helical dichroism can find important applications in acoustic OAM manipulations and chiral sound-matter interactions. 
\end{abstract}

\maketitle

%\tableofcontents

\section{\label{sec:level1}INTRODUCTION}

Chiroptical effects induced by the interaction of light with chiral structures have attracted considerable interest due to broad applications in physics, chemistry, and biology \cite{Valev2013,Wang2014,Ye2017,Hentschel2017,Mun2020}. One intriguing chiroptical effect is circular dichroism (CD), i.e., the differential absorption of right-handed circularly polarized and left-handed circularly polarized lights \cite{berova2000circular}. It has been widely used to detect and analyze chiral structures, including proteins \cite{Greenfield2006}, DNA, and pharmaceutical drugs \cite{garrett2012biochemistry}, and liquid crystals \cite{Binnemans2005}, as well as to control rotatory power \cite{Rogacheva2006}. However, chiral optical interactions in nature are usually very weak due to a significant size mismatch between the chiral structures and light’s wavelength \cite{Mun2020}. Therefore, various artificial structures have been proposed to achieve strong CD effects, including chiral metamaterials \cite{Rogacheva2006,Decker2007,Zhang2009, Decker2009, Zhao2010}, chiral metasurfaces \cite{Kim2014}, and gyroid structures \cite{Saba2011,Oh2013,Dolan2015}. Interestingly, a strong CD effect can also be realized by using an achiral metal sphere excited by a linearly polarized light \cite{2021arXiv210800286J}.  Akin to CD, optical vortex beams (i.e., helical lights) carrying opposite orbital angular momentum (OAM) can also manifest different absorptions when interacting with chiral structures, which is referred to as optical helical dichroism (HD) \cite{Mun2020}. In contrast to CD resulting from the coupling between electric and magnetic dipoles \cite{schaferling2017chiral}, HD is attributed to the electric quadrupole moments induced by the interaction of optical OAM with chiral structures\cite{Forbes2021}, which has been verified theoretically \cite{Babiker2002,Wu2015,Wang2017, Forbes2018} and experimentally \cite{Brullot2016,Ni2021}. 

Despite the extensive study of CD and HD effects in optics, the exploration of their counterparts in acoustics has not been reported yet. Sound propagating in air/fluids is a longitudinal wave carrying no intrinsic spin angular momentum \cite{Bliokh2019,Wang2021}, indicating the absence of an acoustic analogue of optical CD. However, sound can carry intrinsic OAM in the form of acoustic vortex beams (i.e., helical sounds) characterized by a topological charge $q$ \cite{Lu2016,Jiang2016,Esfahlani2017, Liu2020,Fan2020,Fu2020, Zou2020}, which can give rise to many novel phenomena and applications similar to optical OAM, such as acoustic micromanipulations \cite{Demore2012,Baresch2016}, acoustic communications via OAM multiplexing \cite{Shi2017}, acoustic spin-redirection geometric phase \cite{Wang2018}, and acoustic Fedorov-Imbert linear shift\cite{Wang2021a}. An interesting question is that whether sound can have the phenomenon of HD. In this paper, we report that a one-dimensional (1D) periodical lattice of chiral resonators can induce differential absorption of opposite helical sounds (i.e., helical sounds carrying opposite OAM), which we refer to as acoustic HD. This phenomenon is attributed to the interaction of acoustic OAM with the chiral resonators. We show that the chiral resonators with homogenous loss respecting $C_4$ rotation symmetry induce weak acoustic HD. To achieve a large HD, we engineer the loss region to break the $C_4$ rotation symmetry and induce the coupling of opposite chiral dipole states of the resonators. This gives rise to OAM bandgaps and non-Hermitian exceptional points (EPs), which can significantly enhance the acoustic HD. We show that the underlying physics can be well captured by an effective Hamiltonian taking into account the chiral dipole bands of the 1D lattice. These results provide new mechanisms to manipulate acoustic OAM (e.g., selective absorption or transmission of acoustic OAM) and can trigger more explorations of acoustic chiral-matter interactions.

We organize the paper as following. In the Section \ref{sec:level2}, we describe the 1D lattice of chiral resonators and the properties of its band structure and eigenmodes. In Section \ref{sec:level3}, we discuss the acoustic HD effect in two types of the lossy lattice with $C_4$ and $C_2$ symmetry, respectively, where the symmetry is determined by the loss region. The physics for the HD effect are illustrated in Section \ref{sec:level4} using an effective Hamiltonian that can reproduce the band structures of the $C_4$ and $C_2$ systems. To further verify our theory, in Section \ref{sec:level5}, we consider another type of lattice with $C_2$ symmetry determined by the unit cell’s geometry. We draw the conclusion in Section \ref{sec:level6}.

\section{\label{sec:level2}1D LATTICE OF CHIRAL RESONATORS}

\begin{figure}[t!]
\centering
\includegraphics[scale=0.9]{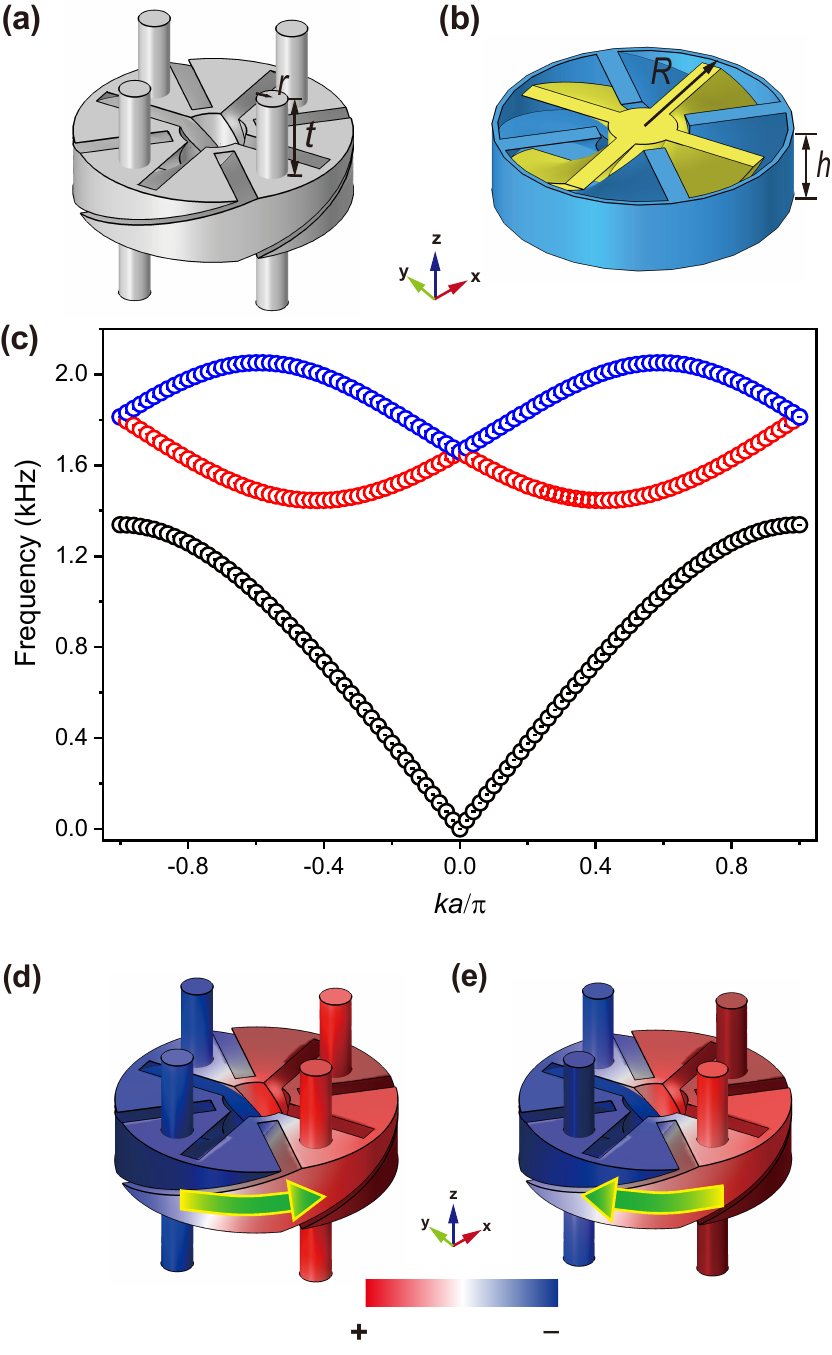}
\caption{ (a) Unit cell of the 1D chiral lattice. (b) A cutaway view of the unit cell. The inner blades are colored in yellow. The outer blades and the shell are colored in blue. The geometric parameters are: $r =0.285$ cm, $t = 1.5$ cm, $R = 2.5$ cm, $h =1.25$ cm. (c) Band structure of the 1D chiral lattice. The eigen pressure fields of the second and third bands at $ka/\pi=0.2$ are shown in (d) and (e), respectively. The green arrows indicate the rotation direction of the pressure fields.} \label{fig1}
\end{figure}

We consider a 1D periodical structure along $z$ direction with the unit cell shown in Figure. \ref{fig1}(a). The unit cell consists of a right-handed chiral resonator (filled with air) with eight tubes that introduce coupling between nearby unit cells. We assume hard boundary conditions at all solid-air interfaces. Figure \ref{fig1}(b) shows a cutaway view of the chiral resonator, where the internal (yellow-colored) and external (blue-colored) blades segmenting the air inside the resonator. Such an internal structure of the resonator enables subwavelength resonances via space coiling. The blades are twisted by $\pi/2$ to introduce chirality into the resonator that breaks the inversion symmetry. The cylindrical resonator has a radius of $R=2.5$ cm and a height of $h=1.25$ cm. The tubes have radii of $r=0.285$ cm and a height of $t=1.5$ cm. 

We conducted full-wave simulations of the periodic structure and computed its band structures by using a finite-element package COMSOL Multiphysics \cite{com}. The results are shown in Fig. \ref{fig1}(c) for the lowest three bands. The first band extending to the static limit corresponds to a monopole mode that has a constant phase of pressure inside the resonator at $k = 0$. The second and third bands correspond to two chiral dipole modes carrying opposite OAM. The two bands are degenerate at $k = 0$ as a result of the $C_4$ rotation symmetry. At $k\neq 0$, the two bands have split due to the inversion symmetry breaking of the chiral resonator. Figure \ref{fig1}(d) and (e) show the eigen pressure fields of the opposite chiral dipole modes at $ka/\pi=0.2$. The blue and red colors denote negative and positive pressures, respectively. The green arrows show the circulating directions of the pressure field as time elapses. Clearly, the modes are of dipole nature and have opposite chirality, i.e., they carry opposite OAM. Importantly, the dipole modes are transverse ( perpendicular to the propagating direction) and give rise to a transverse sound. In the following, we will focus on the interaction of helical sounds with the chiral lattice at the frequencies of the dipole modes and characterize their different absorption properties. 

\section{\label{sec:level3}ACOUSTIC HELICAL DICHROISM}

\begin{figure}[t!]
\centering
\includegraphics[scale=0.9]{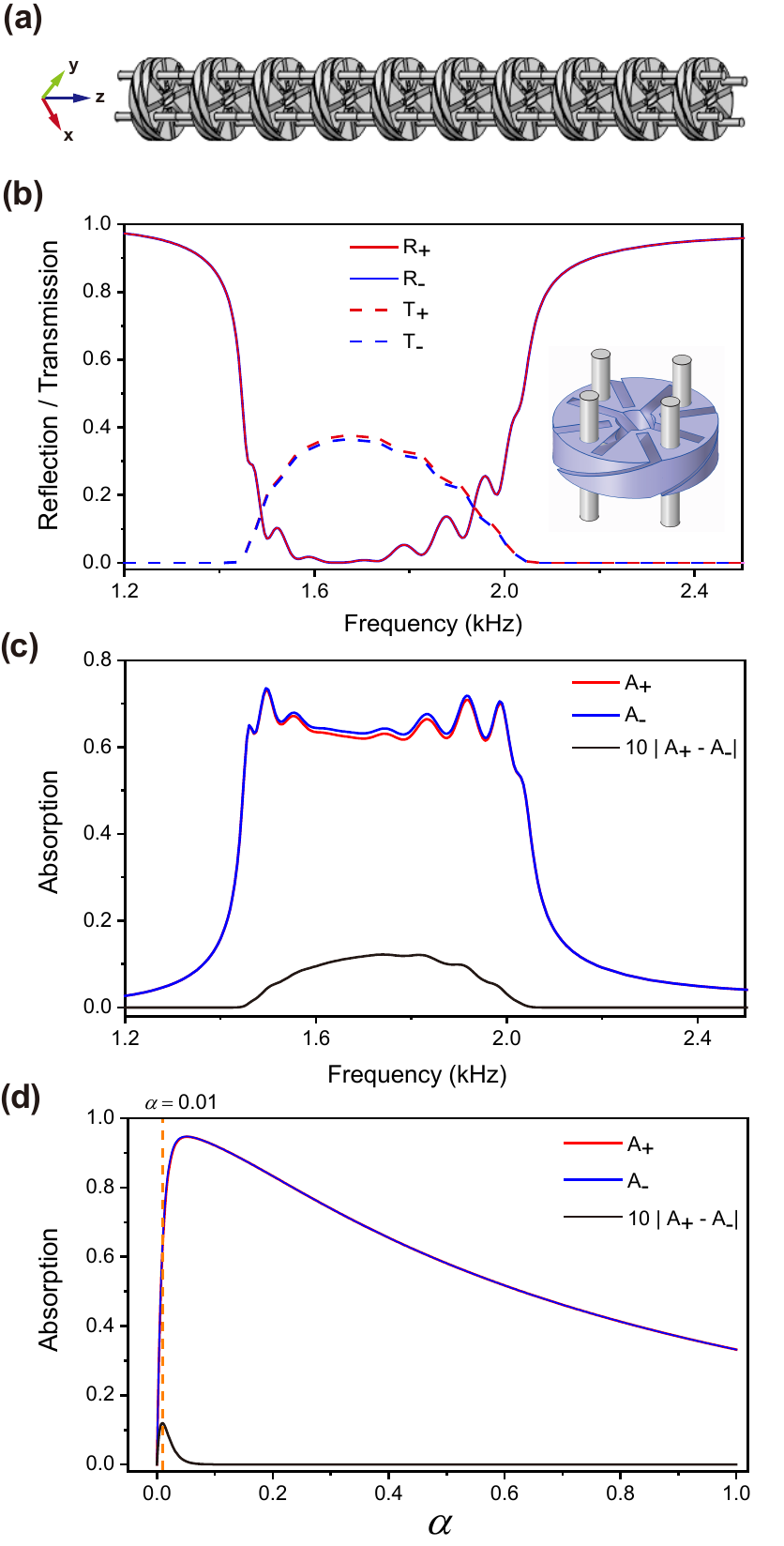}
\caption{ (a) The 1D chiral lattice consisting of 10 unit cells. (b) Transmissions, reflections, and (c) absorptions of helical sounds carrying opposite OAM. The loss is homogeneously added to all the resonators, and the loss parameters $\alpha=0.01$. (d) Absorptions of opposite helical sounds as a function of loss parameters $\alpha$ at $f = 1.8$ kHz. The differential loss is multiplied by ten.} \label{fig2}
\end{figure}

To demonstrate the phenomenon of acoustic HD, we consider the 1D lattice consisting of 10 unit cells shown in Fig. \ref{fig2}(a). We excite the structure via the tubes on the left side of the lattice and calculate the power transmission and reflection coefficients  by using COMSOL. For helical sound with topological charge $q=+1$, the incident pressure at the four input tubes has a phase of $0$, $0.5\pi$, $\pi$ and $1.5\pi$ in the azimuthal direction. For helical sound with a topological charge $q=-1$, the above phases take an extra minus sign. To achieve sound absorption, we introduce loss into the resonators by adding an imaginary part to the speed of sound as $v(1+i\alpha)$, where $v$ is the speed of sound in air and $\alpha$ characterizes the loss strength. Let us denote the reflection and transmission of power as $R_\pm$ and $T_\pm$, respectively, where “$\pm$” denotes the sign of the topological charge $q$ carried by the input helical sounds. Then, the absorption for sound with $q=\pm1$ and the differential absorption charactering the acoustic HD can be determined as:

\begin{align}
    A_\pm&=1-R_\pm-T_\pm, \label{eq:1}\\
    \Delta A&=|A_+-A_-|.
    \label{eq:2}
\end{align}

We first consider the case with homogenous loss introduced into the chiral resonators, as shown by the inset in Figure 2(b), where the blue region of the resonator contains loss. The $C_4$ rotation symmetry is still maintained in this case. The numerical simulation results for the reflection and transmission coefficients  are shown as solid and dashed lines in Fig. \ref{fig2}(b). We have set the loss parameters $\alpha=0.01$. As seen, the reflections of opposite helical sounds, i.e., $R_+$ and $R_-$, are almost identical, while the transmissions $T_+$ and $T_-$ are slightly different. The absorption coefficients obtained using Eq.~(\ref{eq:1}) are shown in Fig. \ref{fig2}(c) as solid blue and red lines. We see that the absorptions of opposite helical sounds have small differences within the frequency range [1.45 kHz, 2.04 kHz], corresponding to the frequencies of the chiral dipole bands in Fig. \ref{fig1}(c). The solid black line shows the differential absorption multiplied by ten, i.e., $\Delta A\times 10$. To understand whether the weak HD is attributed to the small loss parameter $\alpha =0.01$, we calculated the absorption at the frequency of 1.8 kHz while varying the value of $\alpha$, and the results are shown in Fig. \ref{fig2}(d). We see that, as  $\alpha$ is enlarged, the absorptions of opposite helical sounds first increase and then decrease. The same trend is also found for the differential absorption, which has a maximum value $\Delta A\approx 0.01$ at $\alpha=0.01$ (marked by the dashed yellow line). Therefore, this lossy lattice with $C_4$ symmetry cannot induce large HD.

\begin{figure}[t!]
\centering
\includegraphics[scale=0.9]{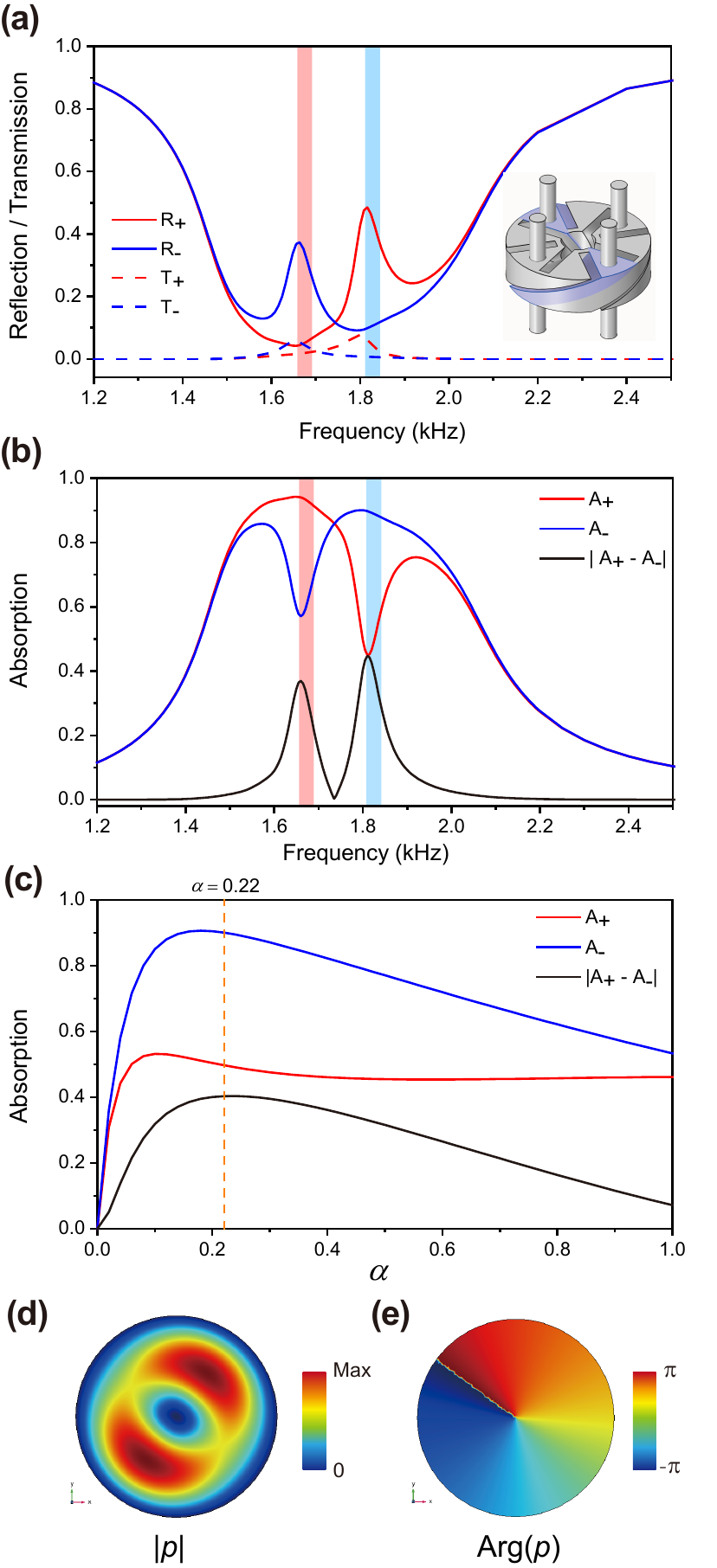}
\caption{ (a) Transmissions, reflections, and (b) absorptions of helical sounds carrying opposite OAM. The loss is selectively added to two opposing sections of the chiral resonators, and we set $\alpha=0.22$. The red and blue ribbons mark the OAM gaps (see main text for definition). (c) Absorption of opposite helical sounds as a function of the loss parameter $\alpha$ at $f =1.8$ kHz. (d) The amplitude and (e) phase of output pressure field under the excitation of  a ``linearly" polarized input sound for $\alpha=0.22$ and $f =1.82$ kHz.} \label{fig3}
\end{figure}

We now consider the lattice with inhomogeneous loss introduced into the chiral resonators, as shown by the inset in Fig. \ref{fig3}(a), where only two sections (blue-colored) of the chiral resonator contain loss. The unit cell now possesses $C_2$ rotation symmetry. We simulated the transmission and reflection of the lattice for helical sounds with topological charge $q=\pm 1$, and the results are shown in Fig. \ref{fig3}(a) as the solid and dashed lines, where the loss parameter is $\alpha=0.22$. We see that both the transmission and reflection of $q=+1$ helical sound have a resonance peak at  $f = 1.81$ kHz, while the transmission and reflection of  $q=-1$ helical sound have a resonance peak at $f = 1.66$ kHz. These peaks lead to a large acoustic HD, as shown in Fig. \ref{fig3}(b) by the solid black line. Remarkably, the HD can reach about 40\% at the resonance frequencies. The solid blue and red lines denote the absorption $A_\pm$ obtained using Eq.~(\ref{eq:1}), which show a dip at corresponding resonance frequencies. We also calculated the absorption as a function of loss strength $\alpha $ at the fixed frequency $f = 1.8$ kHz, and the results are shown in Fig. \ref{fig3}(c). Similar to the $C_4$ system, as $\alpha $ is enlarged, the absorptions of the helical sounds first increase and then decrease. A similar feature is also observed for the differential loss $\Delta A$. The maximum value of $\Delta A$ appears at $\alpha=0.22$, which is the value we set for computing the results in Fig. \ref{fig3}(a) and (b). Apparently, the acoustic HD in this $C_2$ system is much stronger than that of the $C_4$ system in Fig. \ref{fig2}. We will show that the strong acoustic HD can be attributed to the combined effect of OAM bandgaps and non-Hermitian EPs. The different values of $T_+$ and $T_-$ in Fig. \ref{fig3}(a) suggest that the chiral lattice can be employed to generate helical sound with achiral excitation. As a demonstration, we excite the lattice by setting the phase of the pressure at four input tubes to be $0$, $0$, $\pi$, and $\pi$, respectively, corresponding to a “linearly" polarized input sound. The amplitude and phase of the transmitted pressure at $f = 1.81$ kHz are shown in Fig. \ref{fig3}(d) and (e). We notice that the pressure amplitude has an approximate donut shape with zero value in the center, and the phase pattern shows a $2\pi$ variation in the azimuthal direction. These confirm the generation of $q=+1$ helical sound in the transmitted field. The distortion of amplitude donut is attributed to the residue sound with $q=-1$. We note that similar property also exists in the reflected sound, as indicated by the different values of $R_+$ and $R_-$ at the frequencies of the dipole bands in Fig. \ref{fig3}(a). 

\section{\label{sec:level4}COMPLEX BAND STRUCTURES AND EXCEPTIONAL POINTS}    

\begin{figure}[t!]
\centering
\includegraphics[scale=0.9]{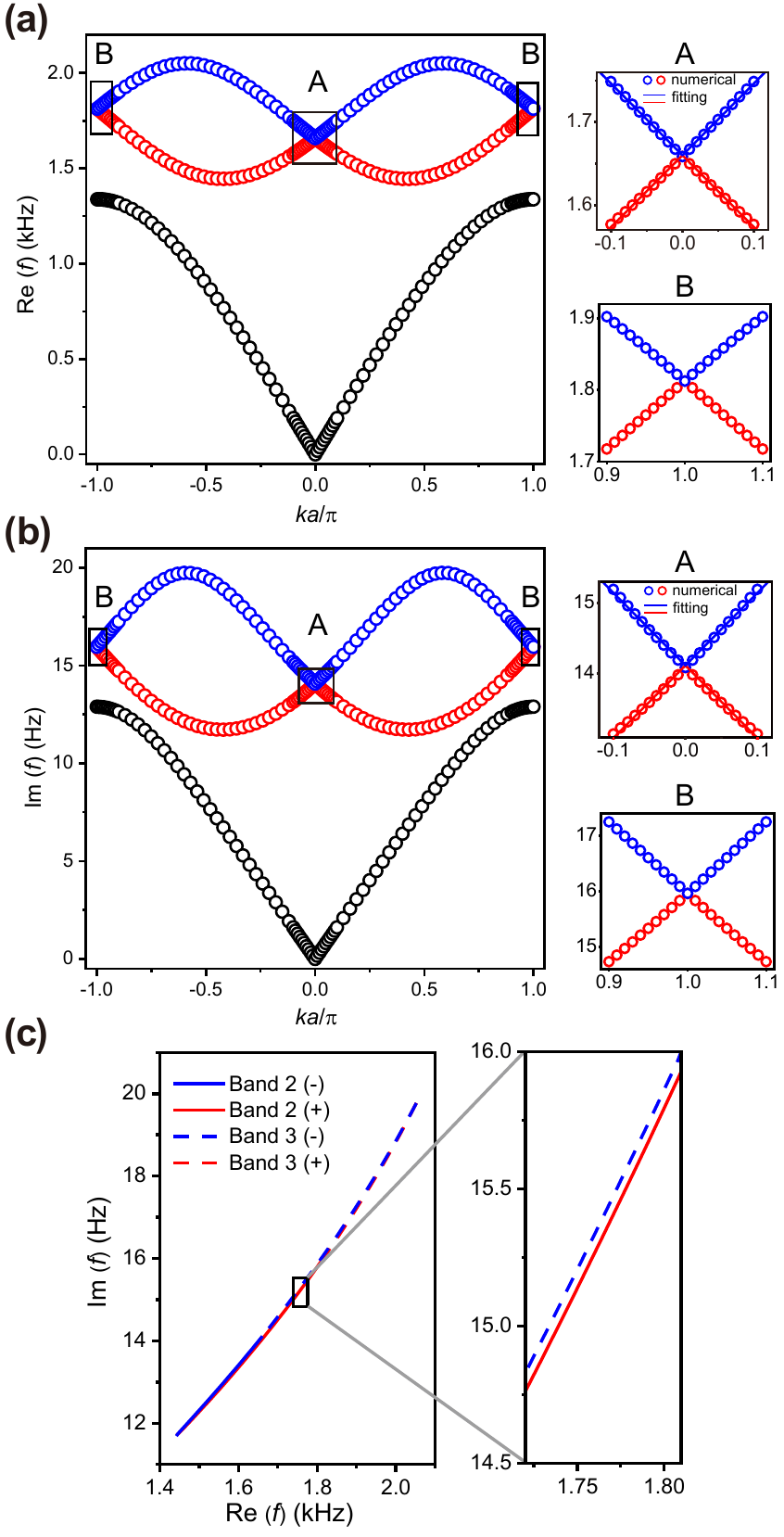}
\caption{ (a) The real part and (b) the imaginary part of the complex band structure for the $C_4$ system with $\alpha=0.01$. The right insets show the zoom-ins of the bands near the zone center and boundaries, as indicated by the black rectangles in (a) and (b). (c) Eigen frequencies corresponding to the second and third bands plotted in the complex plane. The red (blue) color denotes eigenstates with positive (negative) OAM. The right inset is a zoom-in showing the difference of the eigen frequencies.} \label{fig4}
\end{figure}

To understand the underlying physics of the acoustic HD, we study the complex band structures of the 1D lattice with damping. We first consider the lossy lattice with $C_4$ symmetry corresponding to the case of Fig. \ref{fig3}. The numerically computed complex band structures for $\alpha=0.01$ are shown in Fig. \ref{fig4}(a) and (b), respectively. We notice that the real and imaginary parts of eigen frequencies have a similar structure. The imaginary part are positive because of the time-harmonic convention $e^{i\omega t}$ adopted in numerical simulations. The right insets (labelled as A and B) show the zoom-ins of the bands near the zone center and boundaries, as marked by the black rectangles (Inset B corresponds to the composition of two zoom-in regions in Fig. \ref{fig4}(a)).  We see that both the real and imaginary parts are degenerate at the zone center and boundaries, which is protected by the $C_4$ rotation symmetry of the lattice and indicates vanished coupling between the opposite chiral dipole modes. 

To understand the weak HD in Fig. \ref{fig2}, we plot the eigen frequencies of the second and third bands in the complex plane, as shown in Fig. \ref{fig4}(c). The solid (dashed) lines correspond to the second (third) band. The red (blue) color denotes a positive (negative) OAM carried by the corresponding eigenstates (i.e., chiral dipole modes), as labelled by a “$+$” (“$-$”) sign in the figure legend. We notice that the OAM carried by the eigenstates of each band can change the sign because the group and phase velocities can change their relative sign \cite{Wang2021}. The right inset in Fig. \ref{fig4}(c) shows a zoom-in of the complex eigen frequencies, from which we observe a small separation between the red and blue lines. Therefore, under the excitation of input helical sounds within this frequency range, the chiral dipole modes with opposite OAM have different damping. This gives rise to the differential absorption of input helical sounds with opposite OAM, i.e., the acoustic HD.      

\begin{figure}[t!]
\centering
\includegraphics[scale=0.9]{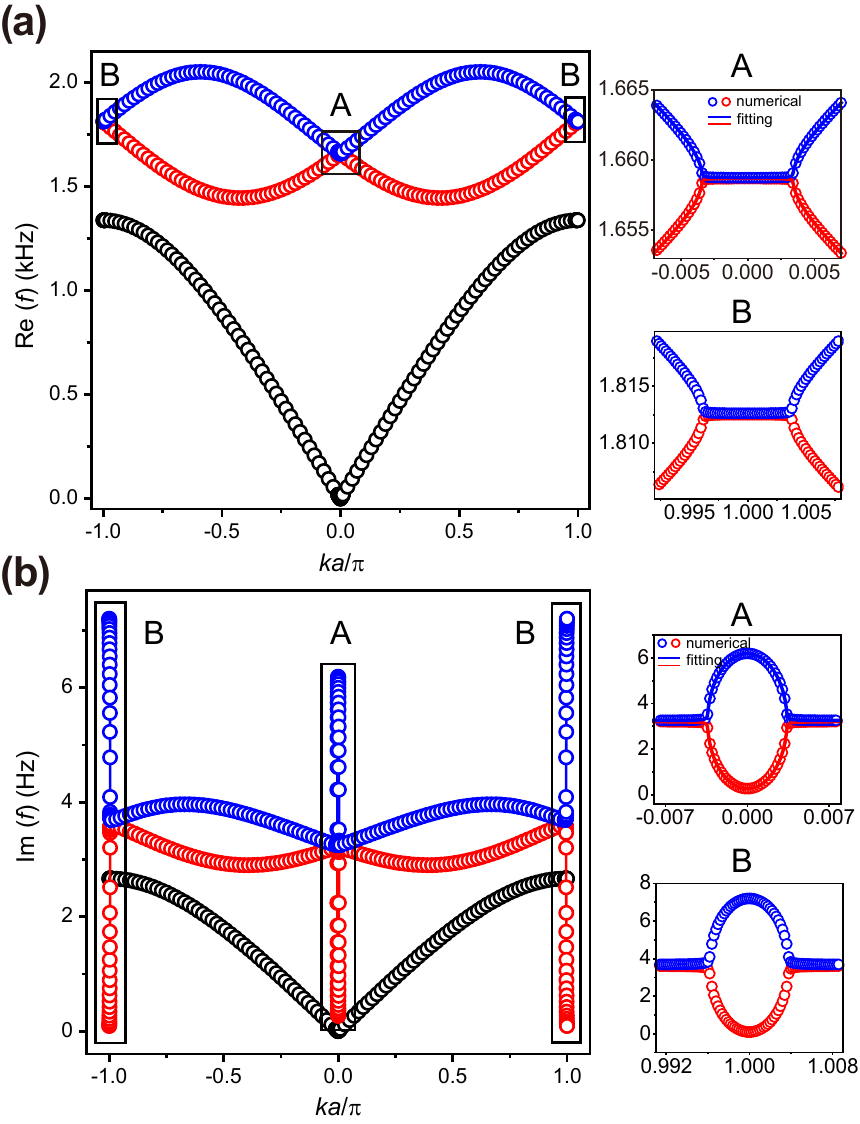}
\caption{ (a) The real part and (b) the imaginary part of the complex band structure for the $C_2$ system with $\alpha=0.01$. The right insets show the zoom-ins of the bands near the zone center and boundaries, as indicated by the black rectangles in (a) and (b). } \label{fig5}
\end{figure}

We now consider the lossy lattice with $C_2$ rotation symmetry shown in Fig. \ref{fig3}(a). Figure \ref{fig5} shows the numerically computed complex band structures for $\alpha=0.01$. In contrast to the $C_4$ system, there are two EPs appearing at the zone center and another two EPs appearing at the zone boundaries. The right insets A and B in Fig. \ref{fig5}(a) and (b) show the zoom-ins of the real and imaginary parts of the bands near the EPs. We notice the typical bifurcation features of EPs: the real parts are degenerate in the symmetry-breaking phase while the imaginary parts bifurcate. These EPs can be considered as derived from the diabolic points in the original lossless system and thus are similar to the EPs emerged in coupled waveguides \cite{Zhang2017,Zhang2018} or EPs spawning from a Dirac point in photonic crystals \cite{Zhen2015}. 

\begin{figure}[t!]
\centering
\includegraphics[scale=0.9]{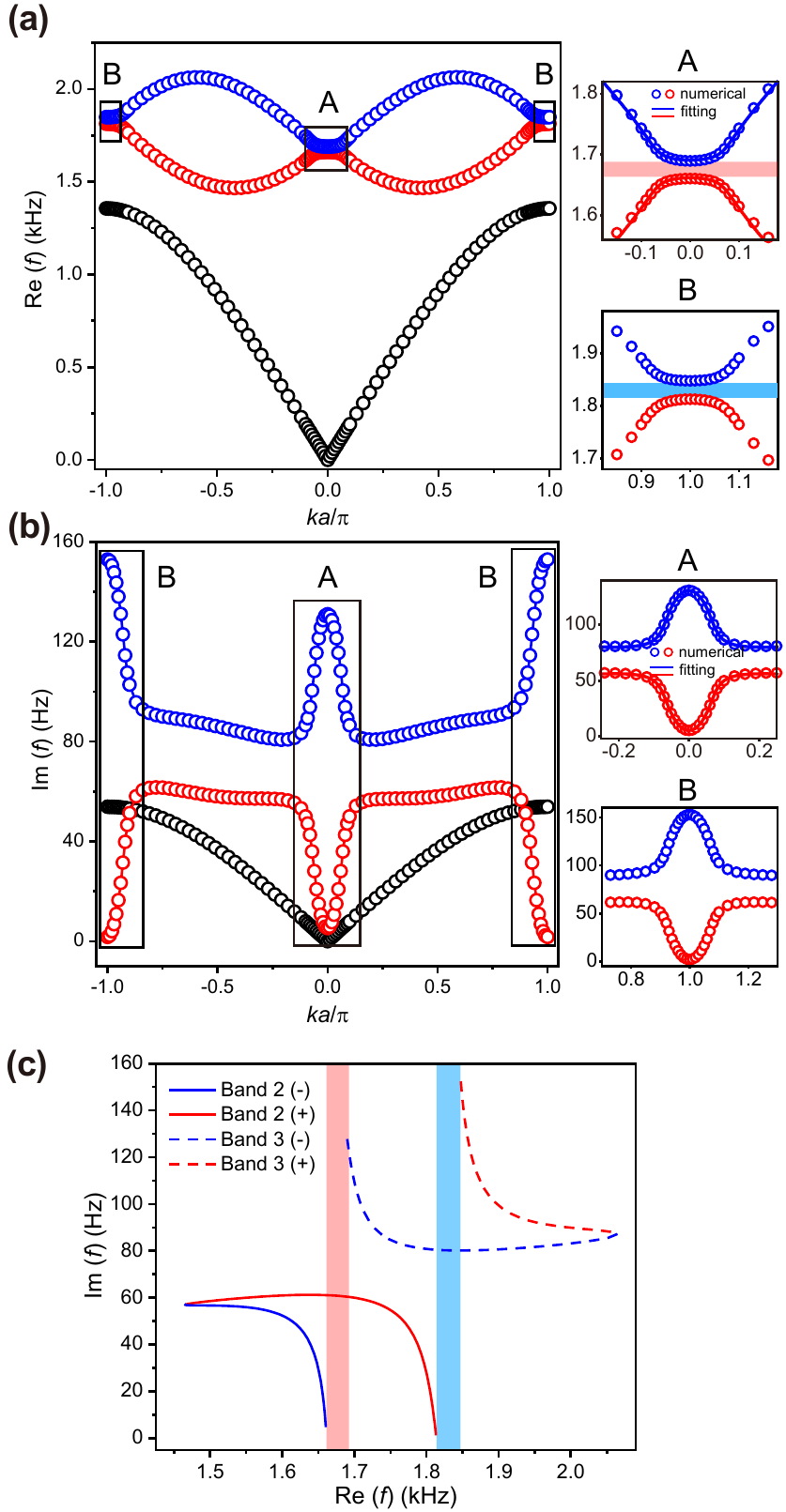}
\caption{ (a) The real part and (b) the imaginary part of the complex band structure for the $C_2$ system with $\alpha=0.22$. The right insets show the zoom-ins of the bands near the zone center and boundaries, as indicated by the black rectangles in (a) and (b). The red (blue) ribbon denotes the OAM gap where the helical sound with positive (negative) OAM can propagate inside the lattice. (c) Eigen frequencies corresponding to the second and third bands plotted in the complex plane. The red (blue) color denotes eigenstates with positive (negative) OAM.} \label{fig6}
\end{figure}

Figure \ref{fig6}(a) and (b) show the complex band structures of the $C_2$ system with a larger loss $\alpha=0.22$, corresponding to the case of Fig. \ref{fig3}(a) and (b). We notice that the general structure of the bands is similar to that in Fig. \ref{fig5}. However, while the general features (e.g., bifurcations) of the EPs can still be observed, there are partial gaps that appear near the zone center and boundaries, making the EPs not well defined. The gaps associated with the real parts of the bands in Fig. \ref{fig6}(a) can be called OAM gaps since the lattice only allows the propagation of one chiral dipole mode within the gaps. Such OAM gaps are similar to the polarization gaps that have been well studied in optical chiral metamaterials \cite{Wu2010}. We apply blue and red colors to label the two gaps in the insets A and B. The frequency ranges of these OAM gaps are also marked in Fig. \ref{fig3}(a) and (b). Noticeably, they agree with the peaks of the acoustic HD. To understand this agreement, we plot the eigen frequencies of the dipole bands in the complex plane, as shown in Fig. \ref{fig6}(c). Similar to Fig. \ref{fig4}(c), we use red (blue) color to denote the eigenstates carrying positive (negative) OAM, as labelled by a symbol of “$+$” (“$-$”) in the figure legend. We also marked the ranges of the two OAM gaps using blue and red ribbons. Remarkably, near the blue/red ribbon, the eigen frequencies of opposite eigenstates have the largest differences in the imaginary parts, which is attributed to the combined effect of the EPs and the OAM gaps. This explains the large acoustic HD in the frequency ranges labelled by the blue and red ribbons in Fig. \ref{fig3}(b).    

We now employ effective Hamiltonians to obtain a better understanding of the physics associated with the EPs \cite{Zhen2015,Miri2019, Wiersig2020, Li2021}. For the homogenous lossy lattice with $C_4$ symmetry, the effective Hamiltonian for the chiral dipole bands at $k\rightarrow0$ can be expressed as: 

\begin{equation}       
H_{C_4}=\left(                 
  \begin{array}{cc}   
    \omega_0-i \gamma & (v_R+i v_I)k\\ 
    (v_R+i v_I)k & \omega_0-i \gamma \\ 
  \end{array}
\right) \label{eq:3}
\end{equation}

\noindent with complex eigenvalues:

\begin{equation}       
\omega_{C_4}=\omega_0 -i \gamma \pm k (v_R+i v_I) , 
\label{eq:4}
\end{equation}

\noindent where $\omega_0$ is the eigen frequency of the chiral dipole modes at  $k=0$ (corresponding to the degeneracy),  $v_R$ and $v_I$ are the real and imaginary parts of the group velocity, $k$ is the wavevector, and $\gamma$ denotes the loss. As for the $C_2$ system, the effective Hamiltonian is

\begin{equation}       
H_{C_2}=\left(                 
  \begin{array}{cc}   
    \omega_0-i \gamma_1 + \frac{a}{2} & (v_R+i v_I)k\\ 
    (v_R+i v_I)k & \omega_0-i\gamma_2-\frac{a}{2} \\ 
  \end{array}
\right) \label{eq:5}
\end{equation}

\noindent with complex eigenvalues:
\begin{equation}      
\begin{split}
\omega_{C_2}=&\omega_0 - i \frac{(\gamma_1 +\gamma_2)}{2}  \\
&\pm\frac{1}{2}\sqrt{[a- i(\gamma_1 - \gamma_2)]^2 - 4k^2 (v_I - i v_R)^2} , 
\label{eq:6}
\end{split}
\end{equation}

\noindent where $\pm a/2 $ describes the gap induced by the symmetry breaking and $\gamma_{1,2}$ denotes the loss of opposite chiral dipole modes at $k=0$. Here, $\gamma_1\ne \gamma_2$ due to the inhomogeneous loss added to the chiral resonator. 

We apply Eqs.~(\ref{eq:4}) and (\ref{eq:6}) to fit the numerically obtained complex band structures in the insets A of Figs. \ref{fig4}-\ref{fig6}, from which we obtain the values for the parameters in the effective Hamiltonian. For the $C_4$ system in Fig. \ref{fig4}(a) and (b), the obtained parameters are   $\gamma=-14.12$ Hz, $v_R=11.71$ m/s, $v_I=0.14$ m/s, and $\omega_0=1660.07$ Hz. For the $C_2$ system in Fig. \ref{fig5}(a) and (b), the obtained parameters are: $a=0.06$ Hz, $\gamma_1=-6.19$ Hz, $\gamma_2=-0.27$ Hz, $v_R=11.81$ m/s, $v_I=0.02$ m/s, and $\omega_0=1658.72$ Hz. For the $C_2$ system in Fig. \ref{fig6}(a) and (b), the obtained parameters are: $a=29.18$ Hz, $\gamma_1=-130.81$Hz, $\gamma_2=-5.51$ Hz, $v_R=11.76$ m/s, $v_I=0.40$ m/s, and $\omega_0=1676.43$ Hz. Here, the loss parameters take negative values due to the time convention $e^{i\omega t}$ adopted in the numerical simulations. The  analytical fitting results are shown as solid blue and red lines in the same insets. As seen, the analytical results given by the effective Hamiltonians well agree with the full-wave numerical results, demonstrating the validity of the effective Hamiltonians. The above analytical model provides a clear physical picture for the stronger acoustic HD in the C2 system. In the $C_4$ system with homogeneous loss, the eigenstates of the two bands at $k=0$ are orthogonal and have no coupling. The dampings of the eigenstates with opposite OAM are approximately equal due to the homogeneous material loss. In the $C_2$ system with inhomogeneous material loss, the symmetry breaking induces coupling and loss difference between the eigenstates of the two bands at $k=0$, leading to the EPs and OAM gaps. The EPs induce bifurcations of the imaginary parts of the eigen frequencies, and the OAM gap enables selective transmission of one eigenstate. These together give rise to the enhanced acoustic HD in comparison with that of the $C_4$ system.

\section{\label{sec:level5}$C_2$ LATTICE WITH HOMOGENOUS LOSS}

\begin{figure*}[t!]
\centering
\includegraphics[scale=0.95]{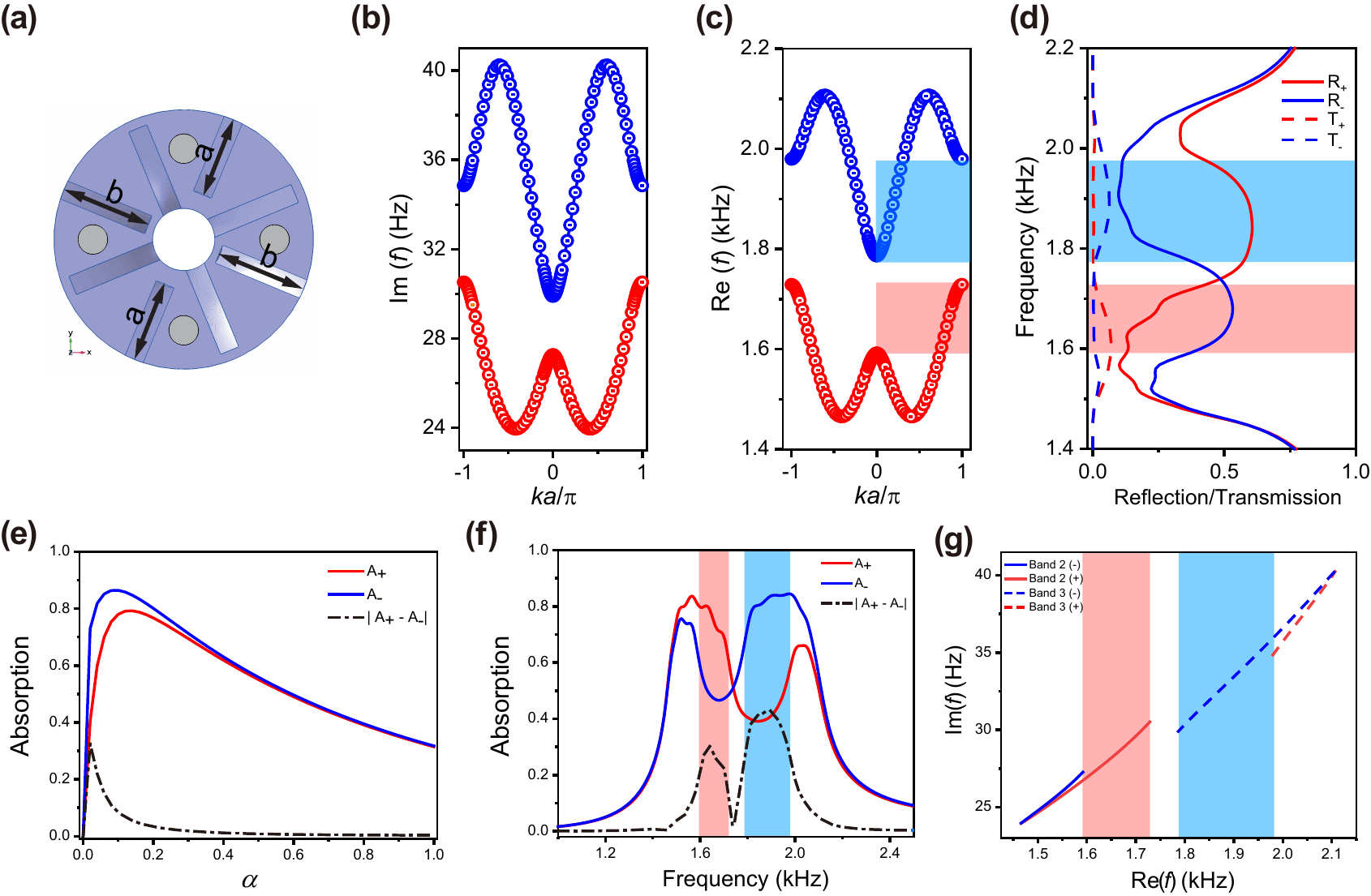}
\caption{ (a) The unit cell with a geometry satisfying $C_2$ rotation symmetry. The lengths of the outer blades are not equal: $a\ne b$ ($a=1.56$ cm, $b=1.76$ cm) (b) The real part and (c) the imaginary part of the complex band structure. (d) The transmission and reflection of helical sounds carrying opposite OAM. We considered the 1D lattice with 10 unit cells and $\alpha=0.02$. (e) The absorption of opposite helical sounds as a function of loss parameter $\alpha$. (f) The absorption of opposite helical sounds corresponding to the case of (d). (g) The complex eigen frequencies of the two dipole bands plotted in the complex plane. The red and blue ribbons in (d), (f), and (g) denote the OAM band gaps.} \label{fig7}
\end{figure*}                                                                       

In the $C_2$ lattice system in Fig. \ref{fig3}, the symmetry breaking is induced by the inhomogeneous loss in the resonator, i.e., the loss is only added to two opposing sections of the resonator. To further understand the effect of the rotation symmetry, we consider another type of $C_2$ system shown in Fig. \ref{fig7}(a), where the loss is homogeneously added to the whole resonator, and the geometry of the resonator has a $C_2$ rotation symmetry since we set internal blades $a\ne b$. Figure \ref{fig7}(b) and (c) show the imaginary and real parts of the complex band structure of this system, respectively. We notice that this system does not give rise to EPs. Because of the $C_2$ symmetry, the chiral dipole bands have two OAM gaps, as indicated by the blue and red ribbons in Fig. \ref{fig7}(c). Figure \ref{fig7}(d) shows the transmission and reflection of opposite helical sounds for loss $\alpha=0.02$. As seen, in the frequency range of the polarization gaps (denoted by blue and red ribbons), the differential reflection and differential transmission reach the maximums. Figure \ref{fig7}(e) shows the absorption of opposite helical sounds as a function of loss strength $\alpha$. Similar to the previous cases, the acoustic HD first increases and then decreases with a maximum at about $\alpha=0.02$.  Figure \ref{fig7}(f) shows the absorptions of opposite helical sounds, where the frequency ranges marked by the blue and red ribbons correspond to that in Fig. \ref{fig7}(c) and (d). As expected, the maximum acoustic HD appears within the two frequency ranges corresponding to the two OAM gaps. For completeness, we also plot the eigen frequencies of the dipole bands in the complex plane, as shown in Fig. \ref{fig7}(g). Similar to the case in Fig. \ref{fig6}(c), at the same value of Re(\emph{f}), the value of Im(\emph{f}) for the eigenstates with opposite OAM are different, indicating different loss of the two states. However, the difference of Im(\emph{f}) for the same Re(\emph{f}) near the boundaries of the OAM gaps is smaller compared with the case in Fig. \ref{fig6}(c), due to the absence of the EPs-induced bifurcations. These results confirm the important effect of structural symmetry on the absorption of helical sounds.  

\section{\label{sec:level6}DISCUSSION AND CONCLUSION}

In optical systems, strong HD can appear even if the materials of chiral structures contain homogenous loss. This is because the chiral structures break inversion symmetry and the induced optical fields are in general different for incident helical lights carrying opposite OAM. Since the absorption strongly depends on the distribution of the fields, HD naturally appears in such chiral structures. Our results in this paper uncover a nontrivial counterpart of optical HD in acoustics. In contrast to the optical HD, the acoustic HD can be very weak in chiral structures with homogeneous material loss. Without the coupling between the opposite chiral dipoles induced in the structures, the absorption of incident helical sounds with opposite OAM is similar. To enhance the HD, one can induce the coupling between opposite chiral dipoles by breaking the $C_4$ rotation symmetry. In our system, this symmetry breaking is realized by selectively adding material loss to the chiral resonators or engineering the resonators’ geometry. The first approach is also benefited from the interesting physics of non-Hermitian EPs, where the bifurcation of complex band structures in the symmetry-breaking phase can also enhance the acoustic HD. We emphasize that although a periodic lattice of the chiral resonators is considered in this study, the acoustic HD can also happen to a single chiral resonator, except that the differential absorption will be smaller. We have applied complex band structures and effective Hamiltonians in $k$ space to explain the observed phenomena. For a single chiral resonator, such explanations are not applicable since the resonator represents an open scattering system. In this case, a microscopic picture based on multipole expansions may be employed to understand its absorption properties, and the associated physics remains to be explored. Experimental demonstration of the proposed acoustic HD is possible. The chiral resonators can be fabricated by using 3D printing and then assembled into a 1D lattice. Loss can be introduced into the lattice by adding absorbing materials (e.g. foams) into the chiral resonators. The incident helical sounds can be generated by using four speakers that couple to the four tubes of the resonator. The absorption can be determined from the measured transmission and reflection of the structure.

In conclusion, we proposed a chiral lattice structure that can selectively absorb helical sounds with opposite OAM. The phenomenon represents the acoustic counterpart of optical HD effect.  We have shown that this acoustic HD strongly depends on the rotation symmetry of the lattice. The structure with $C_2$ symmetry can give rise to a much larger HD compared to the structure with $C_4$ symmetry. This enhancement is attributed to the OAM gaps and non-Hermitian EPs induced by the coupling of opposite chiral dipole modes of the resonators. The results pave the way for further investigations of chiral sound-matter interactions in artificial structures and metamaterials. The proposed structures can also find important applications in manipulations of acoustic OAM.

\begin{acknowledgments}
The work described in this paper was supported by grants from the Research Grants Council of the Hong Kong Special Administrative Region, China (Project No. CityU 21302018) and City University of Hong Kong (No. 9610388). We thank Prof. C. T. Chan and Dr. G. Ma for their useful comments.
\end{acknowledgments}

\nocite{*}

\bibliography{MyCollection}% Produces the bibliography via BibTeX.

\end{document}